\documentclass[12pt]{iopart}
\expandafter\let\csname equation*\endcsname\relax
\expandafter\let\csname endequation*\endcsname\relax
\usepackage{xcolor,diagbox,amsmath,amsfonts,latexsym,amssymb,graphicx,graphics,epsfig,subfigure,color,makeidx}
\usepackage{multirow}
\usepackage[colorlinks,linkcolor=blue,anchorcolor=blue,citecolor=green,urlcolor=blue]{hyperref}
\usepackage[latin1]{inputenc}

\newcommand {\nn}    {\nonumber}
\newcommand {\fc}    {\frac}
\newcommand {\be}    {\begin{equation}}
\newcommand {\ee}    {\end{equation}}
\newcommand {\beq}   {\begin{eqnarray}}
\newcommand {\eeq}   {\end{eqnarray}}
\newcommand {\pa}    {\partial}
\newcommand {\lt}    {\left}
\newcommand {\rt}    {\right}
\begin{document}

\title{Shadow of a charged black hole with scalar hair}

\author{Wen-Di Guo\footnote{guowd@lzu.edu.cn}, Shao-Wen Wei\footnote{weisw@lzu.edu.cn}, and Yu-Xiao Liu\footnote{liuyx@lzu.edu.cn, corresponding author}}
%
%
\address{$^{a}$Lanzhou Center for Theoretical Physics, Key Laboratory of Theoretical Physics of Gansu Province,
School of Physical Science and Technology, Lanzhou University, Lanzhou 730000, People's Republic of China\\
             $^{b}$Institute of Theoretical Physics $\&$ Research Center of Gravitation, Lanzhou University, Lanzhou 730000, People's Republic of China}
\vspace{10pt}
\begin{abstract}
Seeking singularity free solutions are important for further understanding black holes in quantum level. Recently, a five-dimensional singularity free black hole/topological star was constructed [Phys. Rev. Lett. 126, 151101 (2021)]. Through the Kaluza-Klein reduction, an effective four-dimensional static spherically symmetric charged black hole with scalar hair can be obtained. In this paper, we study shadow of this charged black hole with scalar hair in terms of four kinds of observers, i.e., static observers, surrounding observers, freely falling observers, and escaping observers in four-dimensional spacetime. For a spherically symmetric black hole, the shadow is circular for any observer, but the shadow size depends on the motion status of the observer.  On the other hand, the effect of plasma is also investigated by a simple model. The radius of the photon sphere depends on the plasma model. Most importantly, we find that the shadow sizes do not monotonically decrease with $r$ in some cases.
\end{abstract}
%
%
%
%
%
\section{Introduction}

The detection of gravitational waves by LIGO and Virgo collaborations~\cite{Collaboration2016} and the imaging of black hole shadow by Event Horizon Telescope (EHT)~\cite{Collaboration2019a,Collaboration2019,Collaboration2019b,
Collaboration2019c,Collaboration2019d,Collaboration2019e} strengthen our ability to detect the strong gravity regime. This also enhances our ability to test some fundamental physical problems, e.g., do singularities exist~\cite{Berti2015,Cardoso2016,Brahma:2020eos,Bouhmadi-Lopez:2020oia}? Classically, a spacetime singularity locates at $r=0$ for a spherically symmetric black hole. However, from the quantum point view, spacetime should be regular. Some ultra-compact objects such as gravastars~\cite{Mazur2001}, boson stars~\cite{Schunck2008}, wormholes~\cite{Solodukhin2005,Dai:2019mse,Simonetti:2020ivl,Bambi:2021qfo} have been proposed to mimic black holes in the classical description, see Ref.~\cite{Cardoso2019} for a review. However, these objects either need exotic matters or do not have a UV origin. On the other hand, string theory provides us some horizonless models which resemble black holes up to Planck scale above horizon and have smooth microstate geometries, such as fuzz balls~\cite{Gibbons2013}. Usually, these horizonless models need a lot of degrees of freedom in supergravity theories. And it is difficult to relate them to astrophysical observations, e.g., quasinormal modes~\cite{Bena2020} and the deviations from mutlipole moments~\cite{Bena2020a,Bena2020b}. Recently, Ibrahima Bah et al. proposed a five-dimensional topological star/black hole model based on a five-dimensional Einstein-Maxwell theory~\cite{Bah2020,Bah2020a}. In this model, the spacetime is smooth in microstate geometries and similar to the classical black hole in macrostate geometries. So it is interesting to study their observable effects. Actually, the motion of a charged particle in this background has been studied in Ref.~\cite{Lim2021}. Through the Kaluza-Klein reduction, the five-dimensional Einstein-Maxwell theory has been reduced to an effective four-dimensional Einstein-Maxwell-Dilaton theory which possesses a static spherically symmetric solution~\cite{Bena2020a,Bena2020b}. Based on the solution, we can study the observable effects, such as gravitational wave physics and black hole shadow. This will help us to understand the charged black hole better.

We know that, nothing can escape from a black hole in the classical physics, even photons can not. However, due to the strong gravity, the trajectories of photons are curved. So, we can observe photons around the black hole (even that of behind the black hole). Especially, there is a region where photons surround the black hole in unstable circle orbits. This region is usually called the photon sphere. For a Schwarzschild black hole, the photon sphere locates at $r=3M$. In principle photons at this sphere can orbit the black hole forever, but any small perturbation will cause them fall into the black hole or escape to infinity. The shadow size and shape are determined by the photon sphere. The trajectory of photons around a Schwarzschild black hole has been studied ~\cite{Hagihara1931,Darwin1959,Darwin1961,Synge1966,Luminet:1979nyg}. Bardeen investigated the shadow of a rotating Kerr black hole in Ref.~\cite{Bardeen:1973tla}. Usually, the shadow of a spherically symmetric black hole is circular for any observer, and for a rotating black hole the shape will deviate from a sphere. Various of observables have been constructed in order to study the shadow shape and deformation systematically~\cite{Bambi2008,Hioki2009,Bambi2010,Li2013,Tsukamoto2014,Johannsen2015,Abdujabbarov2015,
GhasemiNodehi2015}. Recently, Chang et al., proposed an approach to describe the size and
deformation of shadows using astrophysical observables~\cite{Chang2020,Chang2020a,Chang2021}. This formalism was used to study the shadow of a rotating Hayward-de Sitter black hole~\cite{He2020}.

The first picture of M87* was taken by EHT in 2019~
\cite{Collaboration2019a,Collaboration2019,Collaboration2019b,Collaboration2019c,Collaboration2019d,
Collaboration2019e}. This is the first time that black holes were observed directly. It strengthens the confidence of physicists a lot. With this result, one can study more fine structure near the black hole. Recently, the polarization of the ring and magnetic field structure near the horizon was studied based on the first picture of M87*~\cite{Collaboration2021,Collaboration2021a}. Up to now, black hole shadows and  gravitational lensing have been studied widely~\cite{Wei2015,
Abdolrahimi2015,Oevguen2018,Wang2018,Wei2018,Wei2019,Wang2019,LiuHS2019,Zhu2019,Ding2019,Liu2020,Wei2020,Khodadi2020,
Gralla2019,Mishra:2019trb,Banerjee:2019nnj,Wang2021,Kraniotis:2010gx,Kraniotis:2014paa,Kraniotis:2005zm,Stuchlik:2018qyz,Benavides-Gallego:2018ufb}.

Usually, one considers a black hole in vacuum, for which the photon will orbit the black hole in null geodesics. However, our universe is filled with matters, which will affect the trajectory of photons~\cite{Broderick2003}. So it is important to study the shadow in nonvacuum environment. One of the most common matters in the universe is plasma. It is a dilute medium existing around black holes. For a spherically symmetric black hole, the spherically symmetric plasma only affects the size of the black hole shadow, but for a rotating black hole the shadow shape will also be affected~\cite{Wang2021,Perlick2015,Atamurotov2015,Abdujabbarov2015a,Liu2016,Perlick2017,Yan2019,Dastan2016,Saha2018,Das2019,Babar2020,Fathi2020,Tsupko2021}. Besides, the existence of plasma
might cause superradiant instability~\cite{doi:10.1142/S0217751X13400186,CONLON2018169,Brito2015}. And it will also hinder our ability to test the strong-field gravity~\cite{Cardoso2020}.

In this paper, we will study the shadow of the four-dimensional static spherically symmetric charged black hole with scalar hair observed by four kinds of observers whose motion statuses are static, surrounding the black hole with circular geodesics, freely falling into/escaping from the black hole from/to infinity in the radial direction with and without plasma. We find that without plasma, the shadow size will not monotonically decrease with $r$ for the radial freely falling observer. The existence of plasma will affect the position of the photon sphere and will cause the shadow size smaller. In some cases, the photon sphere will disappear. The nonmonotonically decreasing phenomenon is also found for the case with plasma.

This paper is organized as follows. In Sec.~\ref{themodel}, we give a brief review on the charged black hole with scalar hair. In Sec.~\ref{shadowvaccum}, we calculate the photon sphere of this black hole without plasma. We study the shadow size of the charged black hole with scalar hair in terms of astrophysical observables with four kinds of motion statuses in Sec.~\ref{sizenoplasma}. In Sec.~\ref{effectsofplasma}, we study the effect of plasma through a simple model. Finally, we give the conclusions in Sec.~\ref{conclusion}.

\section{The charged black hole with scalar hair}\label{themodel}

We start with a five-dimensional Einstein-Maxwell theory. The action is given by
\begin{equation}
S=\int d^5x\sqrt{-g}\left(\fc{1}{2\kappa_5^2}R-\fc{1}{4}F^{MN}F_{MN}\right),
\end{equation}
where $\kappa_{5}$ is the five-dimensional gravitational constant and $F$ is the electromagnetic field tensor. Hereafter, we use capital Latin letters $M, N...$ to denote the five-dimensional coordinates, Greek letters $\mu, \nu...$ to denote four-dimensional coordinates. The extra dimension is a warped circle with radius $R_y$. The spherically symmetric metric ansatz is~\cite{Stotyn2011}
\begin{eqnarray}
ds^2=-f_S(r)dt^2+f_B(r)dy^2+\frac{1}{f_S(r)f_B(r)}dr^2+r^2d\theta^2+r^2\sin^2\theta d\phi^2. \label{metric_five}
\end{eqnarray}
With a magnetic flux
\be
F=P\sin\theta d\theta\wedge d\phi,\label{field_strength}
\ee
the solution can be solved as~\cite{Stotyn2011}
\be
f_{B}(r)=1-\frac{r_B}{r},  ~~~~~~~ f_{S}(r)=1-\frac{r_S}{r}, ~~~~~~~  P=\pm\frac{1}{\kappa_{5}^{2}}\sqrt{\frac{3r_Sr_B}{2}}.
\ee
The spacetime has two coordinate singularities located at $r=r_S$ and $r=r_B$, which correspond to a horizon and a degeneracy of the $y$-circle, respectively. The degeneracy of the $y$-circle at $r=r_B$ provides an end to the spacetime. After some coordinate transformations, Ibrahima Bah et al. found that a smooth bubble locates at $r=r_B$~\cite{Bah2020,Bah2020a}. For $r_S\geq r_{B}$, the bubble is hidden behind the horizon. For $r_S<r_{B}$, the horizon cannot be reached because the spacetime ends as the bubble at $r=r_B$. Therefore, for $r_S\geq r_{B}$ and $r_S<r_{B}$, the solution corresponds to a black string and a topological star, respectively~\cite{Bah2020,Bah2020a}.

We can rewrite the metric~\eqref{metric_five} as
\beq
ds_5^2&=&e^{2\Phi}ds_4^2+e^{-4\Phi}dy^2,\\
ds_4^2&=&\hat{g}_{\mu\nu} dx^{\mu} dx^{\nu}
        = f_B^{\frac{1}{2}}\lt(-f_Sdt^2+\frac{dr^2}{f_Bf_S}+r^2d\theta^2+r^2\sin^2\theta d\phi^2\rt)\label{metricfour},
\eeq
where
\be
e^{2\Phi}=f_B^{-1/2},
\ee
and $\Phi$ is a dilaton field. After the Kaluza-Klein reduction, i.e., integrating the extra dimension $y$, the five-dimensional Einstein-Maxwell theory will reduce to a four-dimensional  Einstein-Maxwell-dilaton theory
\be
S_4=\int d^4x\sqrt{-\hat{g}}\left(\fc{1}{2\kappa_4^2}R_4-\fc{3}{\kappa_4^2}\hat{g}^{\mu\nu}\pa_{\mu}\Phi\pa_{\nu}\Phi-\fc{\pi R_y}{2} e^{-2\Phi}\hat{F}_{\mu\nu}\hat{F}^{\mu\nu}\right),
\ee
where $R_4$ is the Ricci scalar constructed by the four-dimensional metric $\hat{g}_{\mu\nu}$. The four-dimensional gravitational constant is $\kappa_4=\fc{\kappa_5}{\sqrt{2\pi R_y}}$. The four-dimensional field strength of the magnetic field can be solved as
\be
\hat{F}=\pm\fc{1}{\kappa_4\sqrt{2\pi R_y}}\sqrt{\fc{3r_B r_S}{2}}\sin\theta d\theta\wedge d\phi.
\ee
For $r_B=0$, the metric~\eqref{metricfour} recovers to the Schwarzschild one.

From the above solution, we can derive the four-dimensional ADM mass $M$ and the magnetic charge $Q_\text{m}$ as
\beq
~~~~~~~~~~~~~~~~~~~~M&=&2\pi\left(\fc{2r_S+r_B}{\kappa_4^2}\right),\nn\\
~~~~~~~~~~~~~~~~~~~~Q_\text{m}&=&\fc{1}{\kappa_4}\sqrt{\fc{3}{2}r_B r_S}.
\eeq
It is also useful to solve $(r_S,r_B)$ for given $(M,Q_\text{m})$. We have two pairs of $(r_S,r_B)$,
\beq
~~~~~r_S^{(1)}&=&\fc{\kappa_4^2}{8\pi}(M-M_{\triangle}),~~~~~r_B^{(1)}=\fc{\kappa_4^2}{4\pi}(M+M_{\triangle}),\label{MQm1}\\
~~~~~r_S^{(2)}&=&\fc{\kappa_4^2}{8\pi}(M+M_{\triangle}),~~~~~r_B^{(2)}=\fc{\kappa_4^2}{4\pi}(M-M_{\triangle}),\label{MQm2}
\eeq
where $M_{\triangle}^2=M^2-\left(\fc{8\pi Q_\text{m}}{\sqrt{3}\kappa_4}\right)^2$. Note that, in five-dimensional spacetime, a smooth bubble locates at $r=r_B$; while in four-dimensional spacetime, when $r<r_B$, $f_B^{1/2}$ becomes imaginary. So, $r=r_B$ is the end of the spacetime. In order to check whether a spacetime is singular, one can compute the curvature invariants, such as Zakhary-McIntosh invariant, Kretschmann scalar and Euler-Poincare invariant~\cite{Kraniotis:2021qah}. We give the expression of the Kretschmann scalar of the four-dimensional effective spacetime considered in this paper
\beq
\hat{K}&=&\hat{R}_{\mu\nu\rho\sigma}\hat{R}^{\mu\nu\rho\sigma}\nn\\
 &=&\frac{3}{64r^7 (r-r_B)^3} \Big[64 r^4 (r_B+2 r_S)^2- 64 r^3r_B \left(2 r_B^2+13 r_B r_S+18 r_S^2\right)\nn\\
 &+&r^2 r_B^2\left(69 r_B^2+872 r_B r_S+2032 r_S^2\right)+477r_B^4 r_S^2\nn\\
 &-&6 r r_B^3 r_S(51 r_B+268 r_S)\Big].
\eeq
It can be seen that the spacetime is singular at $r=r_B$. The metric~\eqref{metricfour} describes a naked singularity ($r_B\geq r_S$) or a black hole ($r_B<r_S$). On the other hand, the Gregory-Laflamme instability~\cite{Gregory:1993vy} will enter the black string scenario. However, the spacetime studied in this paper has a compact extra dimension. This compact extra dimension will lead to a discrete KK mass spectrum which makes it possible to avoid the Gregory-Laflamme instability. Stotyn and Mann demonstrated that if $R_y>\fc{4\sqrt{3}}{3}Q_{\text{m}}$ the solution~\eqref{MQm2} is stable under perturbation, but the solution~\eqref{MQm1} is unstable~\cite{Stotyn2011}. So we only focus on the solution~\eqref{MQm2} in this paper. In the following parts, we only study the case $r_B< r_S$, that is the charged black hole with scalar hair.

There is no observational evidence to exclude the magnetic charge for this solution, so the shadow of this background is worthy to study.

\section{Photon orbits and photon sphere}\label{shadowvaccum}

In this paper, we are interested in shadow of the four-dimensional static spherically symmetric charged black hole with scalar hair. First, we should solve the photon orbits of this black hole. Geometrically speaking, the photon orbits are null geodesics of the spactime. So, we can get the orbits by solving the null geodesic equations. However, we know that the geodesic equations are four coupled second-order differential equations, so it is difficult to solve them directly. Compared with this, Hamiltonian approach is a much easier way. The Hamilton of a photon is given by
\be
H=\fc{1}{2}\hat{g}^{\mu\nu}P_{\mu}P_{\nu},
\ee
where $P^{\mu}=\fc{dx^{\mu}}{d\lambda}$ is the four-momentum of the photon, and $\lambda$ is the affine parameter. With the metric~\eqref{metricfour} and $H=0$, we obtain
\be
-\fc{1}{\sqrt{f_B}f_S}E^2+\sqrt{f_B}f_S(P_r)^2+\fc{1}{\sqrt{f_B}r^2}(P_{\theta})^2+\fc{1}{\sqrt{f_B}r^2\sin^2\theta}L^2=0,\label{Hamiltonian_equation}
\ee
where we have used $P_t=-E$ and $P_{\phi}=L$ with $E$ and $L$ the conserved quantities for the Killing vectors $(\partial_t)^{\mu}$ and $(\partial_{\phi})^{\mu}$, respectively. We can separate the radial part and the angle part of Eq.~\eqref{Hamiltonian_equation} as follows
\beq
~~~~~~~~~~~~~~~-\fc{r^2}{f_S}E^2+f_Bf_Sr^2(P_r)^2&=&-K,\nn\\
~~~~~~~~~~~~~~~~~~~~~~(P_{\theta})^2+\fc{L^2}{\sin^2\theta}&=&K,
\eeq
where $K$ is a constant. Then, we have
\beq
~~~~~~~~~~~~~~~~~~~~(P_{\theta})^2&=&K-\fc{L^2}{\sin^2\theta},\nn\\
~~~~~~~~~~~~~~~~~~~~(P_r)^2&=&\fc{E^2}{f_Bf_S^2}-\fc{K}{f_Bf_Sr^2}.
\eeq
Now, we can write the four-momentum uniformly as
\beq
~~~~~~~~~~~~~~~~~~~~P^t       &=&\fc{E}{\sqrt{f_B}f_S},\nn\\
~~~~~~~~~~~~~~~~~~~~P^r       &=&E\sqrt{1-\fc{f_S}{r^2}\kappa},\nn\\
~~~~~~~~~~~~~~~~~~~~P^{\theta}&=&\fc{E}{f_Br^2}\sqrt{\kappa-\fc{b^2}{\sin^2\theta}},\nn\\
~~~~~~~~~~~~~~~~~~~~P^{\phi}  &=&\fc{Eb}{\sqrt{f_B}r^2\sin^2\theta},
\label{Pofnoplasma}
\eeq
where we have defined $\kappa\equiv\fc{K}{E^2}$ and $b\equiv\fc{L}{E}$. For the null geodesics which can reach infinity, the parameter $b$ is the impact parameter. For large $b$, the light ray can escape from the black hole. While for small $b$, the light ray will fall into the black hole. For the critical case, the photon will orbit to the black hole in a circle forever. The region of these circles is the photon sphere. Note that, the photon sphere is unstable, any perturbation will result in that the photon falls into the black hole or escapes to infinity.

We can always choose the orbit of the photon as the equatorial plane because of the spherical symmetry. In another word, we can choose $\theta=\fc{\pi}{2}$ and $P^{\theta}=0$, and so
\be
\kappa=b^2\label{kappab}.
\ee
The photon sphere is determined by $P_r=0$ and
$\dot{P_r}=0$, where the dot denotes the derivative with respect to the affine parameter $\lambda$. From $P_r=0$ we have
\be
\kappa_{\text{sp}}=\fc{r_{\text{sp}}^2}{{f_S}(r_{\text{sp}})}\label{kappasp}.
\ee
From $\dot{P_r}=0$ and the Hamilton's equation $\dot{P_r}=-\fc{\partial H}{\partial r}$ we can derive
\be
r {f_S}'(r)-2{f_S}(r)=0\label{ps}.
\ee
Solving this equation, the radius of the photon sphere can be obatined as $r_{\text{sp}}=\fc{3}{2}r_S$. This result is similar to that of the Schwarzschild black hole.

\section{shadow size in terms of astrometrical observables}\label{sizenoplasma}

For a spherically symmetric black hole, the shape of the shadow is a sphere, the size of the shadow can be described by the angle between the light rays coming from the photon sphere of the black hole.

We know that, the observed angle of any two light rays $(k^{\mu}, w^{\nu})$ for an observer $u^{\mu}$ is
\be
\cos{\Psi}\equiv\fc{\hat{g}_{\mu\nu} \gamma^{\mu}_{\rho}w^{\rho}\gamma^{\nu}_{\sigma}k^{\sigma}}
{\sqrt{\hat{g}_{\alpha\beta}\gamma^{\alpha}_{\rho}w^{\rho}\gamma^{\beta}_{\sigma}w^{\sigma}}\sqrt{\hat{g}_{\alpha\beta}\gamma^{\alpha}_{\rho}k^{\rho}\gamma^{\beta}_{\sigma}k^{\sigma}}},
\label{cospsi}
\ee
where $\gamma^{\mu}_{\nu}$ is the projector of the observer. That is to say,
\be
\gamma^{\mu}_{\nu}\equiv\delta^{\mu}_{\nu}+u^{\mu}u_{\nu}.
\ee
Substituting this into Eq.~\eqref{cospsi}, we can rewrite the angle as
\be
\cos{\Psi}=\fc{w_{\mu}k^{\mu}}{u_{\alpha}w^{\alpha}u_{\beta}k^{\beta}}+1,
\ee
where we have used $w_{\mu}w^{\mu}=k_{\mu}k^{\mu}=0$ and $u_{\mu}u^{\mu}=-1$. Based on this, Chang and Zhu proposed an approach to describe the shadow of a black hole in terms of astrophysical observables~\cite{Chang2020,Chang2020a,Chang2021}. Using three light ryas from the photon sphere, they defined three angles to describe the size and the deformation of the shadow. For a spherically symmetric black hole, the shape of its shadow is circular for any observer~\cite{Chang2020a}. So we only need to calculate the size of the shadow.

\begin{figure*}[htb]
\begin{center}
\includegraphics[width=5.7cm]{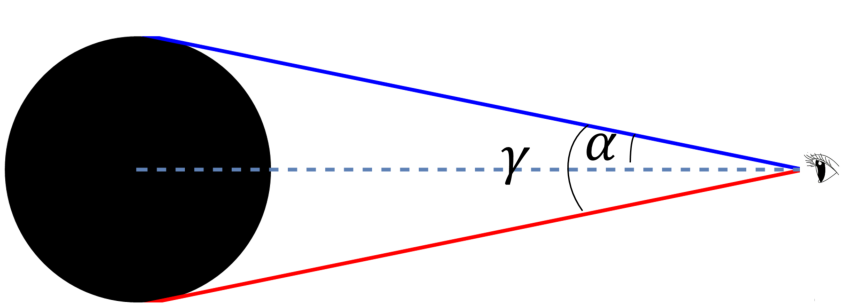}
\end{center}
\caption{Schematic diagram for the angles which are related to the angular diameter of the shadow. The angle $\gamma$ is used in this paper. The angle $\alpha$ is the bending angle in Synge's work~\cite{Synge1966}.}
\label{gamma}
\end{figure*}

For a black hole shadow, the size depends on the two light rays from the photon sphere with opposite angular momenta. The two light rays are described by
\beq
~~~~~~~~~~~~~~~~~~~~k^{\mu}&=&P^{\mu}|_{\kappa=\kappa_{\text{sp}},b=b_{\text{sp}}},\nn\\
~~~~~~~~~~~~~~~~~~~~w^{\mu}&=&P^{\mu}|_{\kappa=\kappa_{\text{sp}},b=-b_{\text{sp}}}.
\eeq
Note that, we have denoted $b_{\text{sp}}=\fc{r_{\text{sp}}}{\sqrt{{f_S}(r_{\text{sp}})}}$. So the angle $\gamma$ which is related the angular diameter is
\be
\cos{\gamma}=\fc{w_{\mu}k^{\mu}}{u_{\alpha}w^{\alpha}u_{\beta}k^{\beta}}+1.\label{angulargamma}
\ee
Here the angle $\gamma$ is twice as large as the bending angle $\alpha$ in Synge's work~\cite{Synge1966}. The schematic diagram for the two angles is shown in Fig.~\ref{gamma}.

Next, we will study the size of the shadow with respect to four kinds of observers whose motion statuses are static, surrounding the black hole with a circular geodesic, freely falling into, and escaping from the black hole in the radial direction, respectively. We will use the subscripts ``st", ``sur",  ``ff", and ``es" to denote quantities of the four kinds of observers, respectively. First, if the motion of the observer can be neglected, then such observer can be viewed as a static observer. Second, if an observer is not far away from the black hole and surrounds the black hole in a circle, then such observer is called a surrounding observer. Third, if an observer is falling into or escaping from the black hole in the radial direction, then such observer is called a freely falling observer or escaping observer.

For the static observer, only the $t$ component of the four-velocity is nonzero. Using the normalization condition $u^{\mu}u_{\mu}=-1$, we have
\be
u_{\text{st}}^t=f_S^{-1/2}f_B^{-1/4}.
\ee
For the observer who is surrounding the black hole with a circular geodesic, solving the geodesic equation $u^{\nu}\nabla_{\nu}u^{\mu}=0$, we obtain
\be
\fc{(u_{\text{sur}}^t)^2}{(u_{\text{sur}}^{\phi})^2}=\fc{4rf_B+r^2f_B'}{f_Sf_B'+2f_Bf_S'}.
\ee
Combining with the normalization condition, we can solve $(u_{\text{sur}}^t)^2$ and $(u_{\text{sur}}^{\phi})^2$ as
\begin{eqnarray}
~~~~~~~~~~~~~~~~~(u_{\text{sur}}^t)^2     &=&\fc{4rf_B+r^2f_B'}{\sqrt{f_B}(4rf_Bf_S-2r^2f_Bf_S')},\nn\\
~~~~~~~~~~~~~~~~~(u_{\text{sur}}^{\phi})^2&=&\fc{f_Sf_B'+2f_Bf_S'}{\sqrt{f_B}(4rf_Bf_S-2r^2f_Bf_S')}.
\end{eqnarray}
For the observer who is freely falling into/escaping from the black hole in the radial direction, using the same method, we have
\beq
~~~~~~~~~~~~~~~~~~~~~~(u_{\text{ff/es}}^t)^2  &=&\fc{1}{f_B{f_S}^2},\nn\\
~~~~~~~~~~~~~~~~~~~~~~(u_{\text{ff/es}}^{r})^2&=&1-\sqrt{f_B}f_S.
\eeq
Note that, the difference between the observer who is escaping from the black hole and the observer who is falling into the black hole freely from infinity is a minus sign in the $r$-component of the four-velocity. For the freely falling observer, $u_{\text{ff}}^{r}=-\sqrt{1-\sqrt{f_B}f_S}$, and for the escaping observer, $u_{\text{es}}^{r}=\sqrt{1-\sqrt{f_B}f_S}$.

Substituting these four-velocities into Eq.~\eqref{angulargamma}, we can obtain the angles $\gamma$ which are related the angular diameters for the four kinds of observers
\beq\label{angulardiameter}
\cos\gamma_{\text{st}}~ &=&1-\fc{\kappa_{\text{sp}}f_S}{r^2}-\fc{b^2_{\text{sp}}f_S}{r^2},\label{cosstatic}\\
\cos\gamma_{\text{sur}}&=&1-\fc{\kappa_{\text{sp}}+b_{\text{sp}}^2}{r^2}\fc{4rf_Bf_S-2r^2f_Bf'_S}{4rf_B+r^2f'_B-b^2_{\text{sp}}(f_Sf'_B+2f_Bf'_S)},\label{cossur}\\
\cos\gamma_{\text{ff}}~~ &=&1-\fc{\kappa_{\text{sp}}+b_{\text{sp}}^2}{\sqrt{f_B}r^2}\lt(\fc{1}{\sqrt{f_B}f_S}+\sqrt{\fc{1}{\sqrt{f_B}}-f_S}\sqrt{\fc{1}{f_S^2}-\fc{\kappa_{\text{sp}}}{f_Sr^2}}\rt)^{-2},\label{cosff}\\
\cos\gamma_{\text{es}}~~ &=&1-\fc{\kappa_{\text{sp}}+b_{\text{sp}}^2}{\sqrt{f_B}r^2}\lt(\fc{1}{\sqrt{f_B}f_S}-\sqrt{\fc{1}{\sqrt{f_B}}-f_S}\sqrt{\fc{1}{f_S^2}-\fc{\kappa_{\text{sp}}}{f_Sr^2}}\rt)^{-2}\label{coses}.
\eeq
Note that, from Eq.~\eqref{kappab} we have $\kappa_{\text{sp}}$ equals to $b^2_{\text{sp}}$, so Eq.~\eqref{cosstatic} can be rewritten as
\be
\cos\gamma_{\text{st}}=1-2\fc{b^2_{\text{sp}}f_S}{r^2}\label{cosstatic2},
\ee
from which we have
\be
\sin^2\lt(\fc{\gamma_{\text{st}}}{2}\rt)=\fc{b^2_{\text{sp}}f_S}{r^2}.
\ee
The bending angle in Synge's work is~\cite{Synge1966}
\be
\sin^2\alpha=\fc{b^2_{\text{sp}}f_S}{r^2}\label{cosstatic3}.
\ee
It means that our result is consistent with Synge's since $\gamma/2=\alpha$.

\begin{figure*}[htb]
\begin{center}
\includegraphics[width=5.7cm]{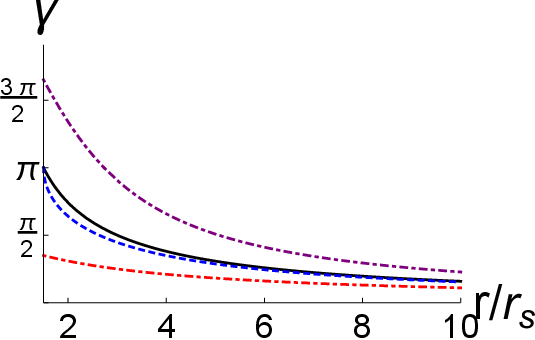}
\end{center}
\caption{The angle $\gamma$ as a function of the radial distance $r/r_S$ for four kinds of observers. The black line, blue dashed line, red dot dashed line, and purple dot dashed line correspond to the static observer, surrounding observer, freely falling observer, and escaping observer, respectively. The parameter $r_B$ is set to $r_B=0.5 r_S$.}
\label{TSnop3os}
\end{figure*}

\begin{figure*}[htb]
\begin{center}
\subfigure[$ $]  {\label{TSnopSR}
\includegraphics[width=5.7cm]{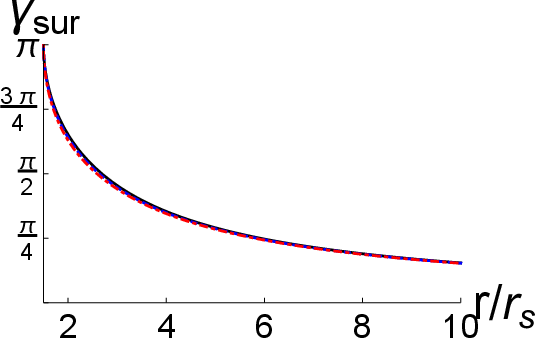}}
\subfigure[$ $]  {\label{TSnopRG}
\includegraphics[width=5.7cm]{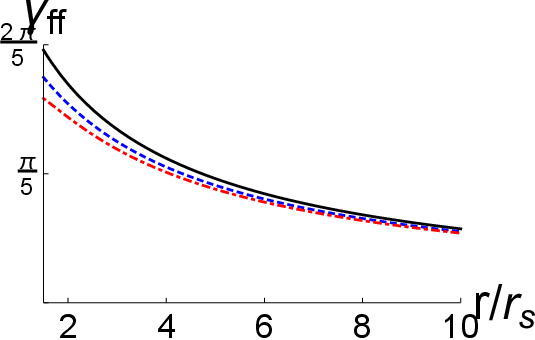}}
\subfigure[$ $]  {\label{TSnopES}
\includegraphics[width=5.7cm]{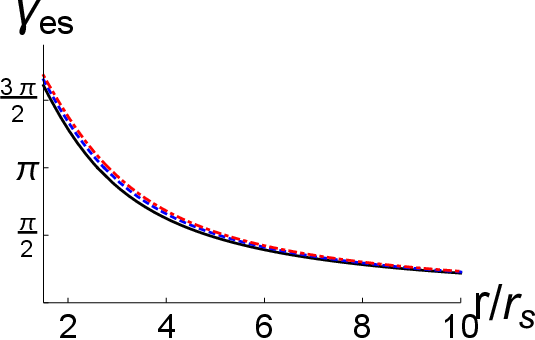}}
\end{center}
\caption{The angle $\gamma$ as a function of the radial distance $r/r_S$. The parameter $r_B$ is set to $r_B=0$ (the black solid lines), $r_B=0.5 r_S$ (the blue dashed lines), and $r_B=0.8 r_S$ (the red dot dashed lines). (a) The surrounding observer. (b) The freely falling observer. (c) The escaping observer.}
\label{surffes}
\end{figure*}

From Eq.~\eqref{angulardiameter} we can see that, for the static observer, the angle $\gamma$ does not depend on the parameter $r_B$; but for the other two cases the angle $\gamma$ depend on $r_B$. That is to say, this charged black hole with scalar hair can not be distinguished from the Schwarzschild black hole by the shadow for the static observer. We plot the angle $\gamma$ for four kinds of observers in Fig.~\ref{TSnop3os}. The effects of the parameter $r_B$ on the angle $\gamma$ for the surrounding observer, freely falling observer, and escaping observer are plotted in Fig.~\ref{TSnopSR}, Fig.~\ref{TSnopRG}, and Fig.~\ref{TSnopES}, respectively. From Fig.~\ref{TSnop3os} we know that, the relation of the angle $\gamma$ for the four kinds of observers is $\gamma_{\text{es}}>\gamma_{\text{st}}>\gamma_{\text{sur}}>\gamma_{\text{ff}}$. Note that, Chang and Zhu concluded that the shadow size tends to be shrunk for a moving observer~\cite{Chang2020a}, but our results shows that, the angular diameter of the shadow for the escaping observer is larger than that of for the static observer. As for the effect of the parameter $r_B$, we see that the larger $r_B$, the smaller the angular diameter for the static observer, surrounding observer, and freely falling observer. But for the escaping observer, the shadow size increases with $r_B$ which can be seen from Fig.~\ref{TSnopES}.

\begin{figure*}[htb]
\begin{center}
\includegraphics[width=5.7cm]{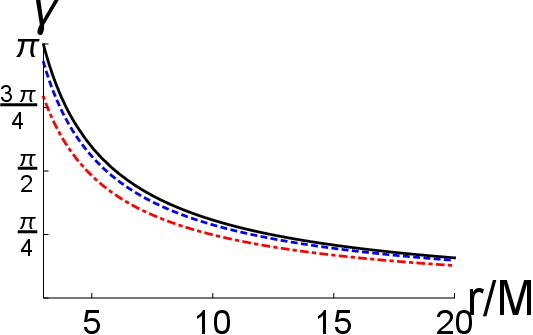}
\end{center}
\caption{The angle $\gamma$ as a function of the radial distance $r/M$ for the static observer. The magnetic charge $Q_\text{m}$ is set to $Q_\text{m}=0.1\bar{M}$ (the black solid line), $Q_\text{m}=0.5\bar{M}$ (the blue dashed line), and $Q_\text{m}=0.8\bar{M}$ (the red dot dashed line), and $\bar{M}$ is defined as $\fc{\sqrt{3}\kappa_4}{8\pi}M$.}
\label{TSstqm}
\end{figure*}

In order to study the effect of the magnetic charge, we should use the mass $M$ and magnetic charge $Q_\text{m}$ to replace $r_B$ and $r_S$ of the angle $\gamma$ in Eqs.~\eqref{cosstatic}-\eqref{coses}. Because the solution~\eqref{MQm1} is unstable under perturbation. So we study the effect of the magnetic charge $Q_\text{m}$ with the solution \eqref{MQm2}. We find that, the shadow size decreases with the magnetic charge $Q_\text{m}$. The plot of the shadow for the static observer is shown in Fig.~\ref{TSstqm}. This is reasonable, because when we fix the mass $M$, and increase the magnetic charge $Q_\text{m}$, the parameter $r_S$ will decrease and $r_B$ will increase. So the change of the shadow is similar to the case when we fix $r_S$ and increase $r_B$. The plot of the shadow for the static observer is shown in Fig.~\ref{TSstqm}, and the situations for the other three kinds of observers are similar to the case of static observer.

\section{Effect of plasma on the shadow}\label{effectsofplasma}

We know that our universe is not vacuum. Instead, it is filled with plasma, a dilute medium which will affect the trajectories of photons. So, it is important to study the effect of plasma on the shadow. Perlick and Tsupko et al. studied the influence of plasma on the shadow of a general spherically symmetric black hole, and it was generalized to an arbitrary transparent dispersive medium case~\cite{Perlick2015,Tsupko2021}.

In this paper, we focus on a nonmagnetized cold plasma. The frequency $\omega_\text{P}$ of the electron plasma only depends on the radial coordinate:
\be
\omega_\text{P}^2(r)=\fc{4\pi e^2}{m}N(r)\label{omegap},
\ee
where $e$, $m$, and $N(r)$ are the electron charge, electron mass, and electron number density of the plasma, respectively. With this, the Hamiltonian of light rays in the plasma can be derived from the the Maxwell's equations~\cite{doi:10.1098/rspa.1980.0040,doi:10.1098/rspa.1981.0011},
\be
H=\fc{1}{2}(\hat{g}^{\mu\nu}P_{\mu}P_{\nu}+\omega_{\text{P}}^2)\label{hamiltonian}.
\ee
Volker Perlick studied a two-fluid plasma model with vanishing pressure in the non-magnetized plasma medium~\cite{Perlick2000}. The author ignored the effect of the electromagnetic wave on the ions, and found that the eikonal equation (the characteristic equation of the system of evolution equations) leads to three Hamiltonians. But only the Hamiltonian~\eqref{hamiltonian} can lead to the correct transverse modes. The light rays solved from the Hamiltonian~\eqref{hamiltonian} is the time-like geodesics of the metric
$\omega_{\text{P}}^2 \hat{g}^{\mu\nu}$ which is conformally equivalent to $\hat{g}^{\mu\nu}$. This can be seen from the fact that when the Hamiltonian changes as follows
\beq
~~~~~~~~~~~~~~~~~\bar{H}=\fc{1}{\omega_{\text{P}}^2}H=\fc{1}{2}(\fc{1}{\omega_{\text{P}}^2}\hat{g}^{\mu\nu}P_{\mu}P_{\nu}+1),
\eeq
the solutions solve from the Hamiltonian equations $\frac{dx^{\mu}}{d\lambda}=\frac{\partial H}{\partial p_{\mu}},~\frac{dp_{\mu}}{d\lambda}=-\frac{\partial H}{\partial x^{\mu}},~ H(x,p)=0$ do not change~\cite{Perlick2000}.
Separating the radial part and the angle part and using $H=0$, we can get that
\beq
~~~~~-\fc{r^2}{f_S}E^2+f_Bf_Sr^2(P_r)^2+\sqrt{f_B}r^2\omega_\text{P}^2&=&-K,\\
~~~~~~~~~~~~~~~~~~~~~~~~~~~~~~(P_{\theta})^2+\fc{L^2}{\sin^2\theta}&=&K.\label{kappa}
\eeq
The four momentum $P^{\mu}$ can be derived in the same procedure as the previous part
\beq
~~~~~~~~~~~~~~~~~~~~~~~~~P^t       &=&\fc{E}{\sqrt{f_B}f_S},\nn\\
~~~~~~~~~~~~~~~~~~~~~~~~~P^r       &=&E\sqrt{1-\fc{f_S}{r^2}\kappa-\sqrt{f_B}f_S\omega_\text{P}^2},\nn\\
~~~~~~~~~~~~~~~~~~~~~~~~~P^{\theta}&=&\fc{E}{f_Br^2}\sqrt{\kappa-\fc{b^2}{\sin^2\theta}},\nn\\
~~~~~~~~~~~~~~~~~~~~~~~~~P^{\phi}  &=&\fc{b}{\sqrt{f_B}r^2\sin^2\theta}.
\eeq
Compared with Eq.~\eqref{Pofnoplasma}, only the $r$ component is affected by the plasma. Due to the presence of plasma,
the tangent vectors of light rays are no longer null. So the expression of the angle $\gamma$ is changed to
\be
\cos{\gamma}=\fc{w_{\mu}k^{\mu}+u_{\mu}u_{\nu}w^{\mu}k^{\nu}}{\sqrt{(u_{\alpha}w^{\alpha})^2-\omega_\text{P}(r)^2}\sqrt{(u_{\beta}k^{\beta})^2-\omega_\text{P}(r)^2}}.\label{cosgammawithp}
\ee

As in the previous section, we choose the orbit of the photon as the equatorial plane, which means $\theta=\fc{\pi}{2}$, $P^{\theta}=0$, and $\kappa=b^2$. The photon sphere is also determined by $P_r=0$ and $\dot{P_r}=0$. But the situation is more complicated here. Form $P_r=0$ we can derive
\be
\kappa_{\text{sp}}=\fc{r_{\text{sp}}^2}{f_S(r_{\text{sp}})}-\sqrt{f_B(r_{\text{sp}})}r_{\text{sp}}^2\fc{\omega_\text{P}^2(r_{\text{sp}})}{E^2}.
\ee
The condition $\dot{P_r}=0$ gives
\be
\fc{f_S'}{f_S^2\sqrt{f_B}}-\fc{2}{f_S\sqrt{f_B}r}+\lt(\fc{f_B'}{2f_B}+\fc{2}{r}\rt)\fc{\omega_\text{P}^2}{E^2}+\lt(\fc{\omega_\text{P}^2}{E^2}\rt)'=0.\label{photonsphereplasma}
\ee
We can solve the radius of the photon sphere from this equation, which obviously depends on the frequency of the plasma.

The angle $\gamma$ for the four kinds of observers can be written as
\beq\label{angulardiameterwithp}
\cos\gamma_{\text{st}}~ &=&1-\lt(\fc{\kappa_{\text{sp}}f_S}{r^2}+\fc{b^2_{\text{sp}}f_S}{r^2}\rt)\fc{1}{1-\sqrt{f_B}f_S\fc{\omega_\text{P}^2}{E^2}},\\
\cos\gamma_{\text{sur}}&=&\fc{-2 f_B^{3/2} r^2 \left(r f_S'-2 f_S\right)\fc{\omega_\text{p}^2}{E^2}+r f_B' \left(f_S b _{\text{sp}}^2-r^2\right)-2 f_B r \left(f_S' \kappa _{\text{sp}}+2 r\right)}{f_B^{3/2} r^2 \left(r f_S'-2 f_S\right)\sqrt{G^2-F^2}}\nn\\
&+&\fc{4 f_B f_S\left(\kappa _{\text{sp}}+b_{\text{sp}}^2\right)}{f_B^{3/2} r^2 \left(r f_S'-2 f_S\right)\sqrt{G^2-F^2}},\\
\cos\gamma_{\text{ff}}~~ &=&1-\fc{1}{r^2}
\fc{\sqrt{f_B}\lt(\kappa_{\text{sp}}+b_{\text{sp}}^2\rt)}{\lt(\fc{1}{f_S}+\sqrt{1-\sqrt{f_B}f_S}\sqrt{\fc{1}{f_S^2}-\fc{\kappa_{\text{sp}}}{f_Sr^2}-\fc{\sqrt{f_B}\omega_\text{P}^2}{E^2f_S}}\rt)^2-\fc{f_B\omega_\text{P}^2}{E^2}},\\
\cos\gamma_{\text{es}}~~ &=&1-\fc{1}{r^2}
\fc{\sqrt{f_B}\lt(\kappa_{\text{sp}}+b_{\text{sp}}^2\rt)}{\lt(\fc{1}{f_S}-\sqrt{1-\sqrt{f_B}f_S}\sqrt{\fc{1}{f_S^2}-\fc{\kappa_{\text{sp}}}{f_Sr^2}-\fc{\sqrt{f_B}\omega_\text{P}^2}{E^2f_S}}\rt)^2-\fc{f_B\omega_\text{P}^2}{E^2}},
\eeq
where $F$ and $G$ are defined as
\beq
~~~~~~~~~~F&=&\fc{2b _{\text{sp}}\sqrt{\lt(f_S f_B'+2 f_B f_S'\rt)\lt(r f_B'+4 f_B\rt)}}{f_B^{3/2} r \lt(2 f_S-r f_S'\rt)}-\fc{\omega_\text{p}^2}{E^2},\nn\\
~~~~~~~~~~G&=&\fc{2 f_B \lt(f_S' b _{\text{sp}}^2+2 r\rt)+f_B' \lt(f_S b _{\text{sp}}^2+r^2\rt)}{f_B^{3/2} r \lt(2 f_S -r f_S'\rt)}.
\eeq

Next, we consider a specific model to study the effect of plasma. For simplicity, we consider the spherically symmetric nonmagnetized pressureless plasma around the black hole~\cite{Perlick2015}. Due to gravity of the black hole, the plasma will fall into the black hole in the radial direction freely. From Eq.~\eqref{omegap} we know that, we need the number density of electrons in the plasma to get the plasma frequency.

We start with the continue equation
\be
\partial_{\mu}(\sqrt{-\hat{g}}\rho u^{\mu})=0,
\ee
where $\rho$ and $u^{\mu}$ are the rest mass density and the four-velocity of the plasma. Here, we consider the plasma is consist of neutral hydrogen, and the four-velocity of the electrons is same to the hydrogen. Because the mass of an electron is negligible compared to the mass of a proton, so the rest mass density $\rho$ is
\be
\rho=m_\text{p} N,\label{rhompN}
\ee
where $m_\text{p}$ and $N$ are the rest mass of a proton and the number density of the protons. Because the plasma is neutral, the number of the electrons is the same to the number of the protons. Due to the spherical symmetry and the stationary accretion, the continue equation becomes
\be
\fc{d(f_Br^2\rho u^r)}{dr}=0.
\ee
Integrating this equation, we obtain that
\be
f_B r^2 \rho u^r=-C,
\ee
where $C$ is an integral constant. In our case, it denotes the mass flux of the plasma. We assume that the plasma is falling into the black hole freely, so the trajectories are radial geodesics. For our background metric, the $r$ component of the four-velocity is
\be
u^r=-\sqrt{1-\sqrt{f_B}f_S},
\ee
With this, we can calculate the mass density as
\be
\rho=\fc{C}{f_Br^2\sqrt{1-\sqrt{f_B}f_S}}.
\ee
Substituting this equation into Eqs.~\eqref{omegap} and~\eqref{rhompN}, we can write the plasma frequency as
\be
\fc{\omega_\text{P}^2}{E^2}=\fc{4\pi e^2\rho}{m_em_pE^2}=\beta\fc{r_S^2}{r^2f_B\sqrt{1-\sqrt{f_B}f_S}},
\ee
where
\be
\beta=\fc{e^2C}{m_em_pE^2r_S^2}.
\ee
For this plasma model, we can only solve the radius of the photon sphere numerically since Eq.~\eqref{photonsphereplasma}
is a higher degree equation of $r$. We show the result for some values of $r_B$ and $\beta$ in Table~\ref{photonspher}. From this table we can see that both the two parameters $\beta$ and $r_B$ have an effect on the radius of the photon sphere. For smaller $r_B$ (the two upper lines in Table~\ref{photonspher}), the value of $r_{\text{sp}}$ increases with $\beta$. But for larger $r_B$ (the other four lines in Table~\ref{photonspher}), the value of $r_{\text{sp}}$ decreases with $\beta$. Besides, when $r_B$ is much larger, $r_{\text{sp}}$ will be smaller than $r_B$. However, $r_B$ is the end of the spacetime, so this result is unphysical. That is, for larger $r_B$, the existence of plasma will result in that the photon sphere
disappears. Besides, from Eq.~\eqref{kappa} and the definition of $\kappa$ we know that $\kappa$ is nonnegative.
This gives an upper bound of the parameter $\beta$, which is listed in Table~\ref{boundofbeta}. This table shows that the upper bound of the parameter $\beta$ decreases with $r_B$.

\begin{table}[!htb]
\begin{center}
\begin{tabular}{|l|c| c| c| c| c |c||}
\hline
\diagbox{$r_B/r_S$}{$r_{\text{sp}}/r_S$}{$\beta$}&~~~~1~~~~&~~~~~2~~~~~&~~~~3 & ~~~~4~~~~  \\
\hline
~~~0.5                        &  1.50942    &           1.51958&  1.53058    &    1.54254   \\
\hline
~~~0.6                         &  1.50418   &           1.50867      &  1.51350     &    1.51872   \\
\hline
~~~0.7                        &  1.49725    &           1.49434&  1.49120   &    1.48783    \\
\hline
~~~0.8                      &  1.48769   &           1.47448& 1.46031& 1.44518\\
\hline
~~~0.9                         &  1.47355    &           1.44450& 1.41300& \\
\hline
~~~1.0                       &  1.45000    &           1.39040& 1.32009& \\
\hline
\end{tabular}
\caption{The radii of the photon sphere with plasma for different values of $r_B$ and $\beta$.}
\label{photonspher}
\end{center}
\end{table}

\begin{table}[!htb]
\begin{center}
\begin{tabular}{|c| c|c|c|c|c|c|}
\hline
$r_B/r_S$               &   0.5   &   0.6   &   0.7   &   0.8   &   0.9   &   1.0  \\
\hline
upper bound of $\beta$  & 4.68879 & 4.50092 & 4.28685 & 4.03709 & 3.73250 & 3.31718 \\
\hline
\end{tabular}
\caption{The upper bound of the parameter $\beta$ for different values of $r_B$.}
\label{boundofbeta}
\end{center}
\end{table}

\begin{figure*}[htb]
\begin{center}
\subfigure[$ $]  {\label{TSP3os}
\includegraphics[width=5.7cm]{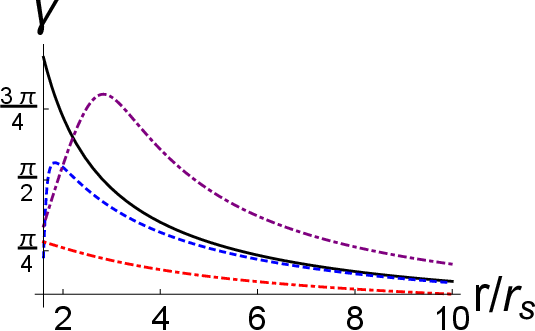}}
\subfigure[$ $]  {\label{TSPbeta3os}
\includegraphics[width=5.7cm]{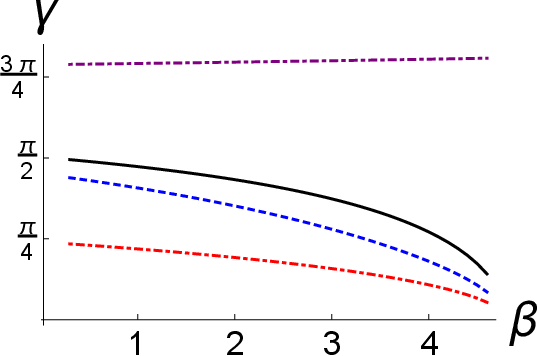}}
\end{center}
\caption{The angle $\gamma$ as a function of the radial distance $r/r_S$ of the observer or the parameter $\beta$ in the plasma model with $r_B=0.5 r_S$. The black solid lines, blue dashed lines, red dot dashed lines, and purple dot dashed lines correspond to the static observer, surrounding observer, freely falling observer, and escaping observer, respectively. (a) The angle $\gamma$ as a function of the radial distance $r/r_S$ for four kinds of observers with $\beta=1$. (b) The angle $\gamma$ as a function of the parameter $\beta$ with the observers located at $r=3r_S$.}
\label{TSP}
\end{figure*}

\begin{figure*}[htb]
\begin{center}
\subfigure[$ $]  {\label{TSPb001}
\includegraphics[width=5.7cm]{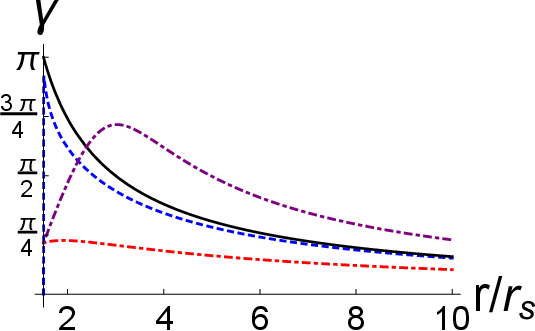}}
\end{center}
\caption{The angle $\gamma$ as a function of the radial distance $r/r_S$ for four kinds of observers in the plasma model with $\beta=0.01$. The black solid lines, blue dashed lines, red dot dashed lines, and purple dot dashed lines correspond to the static observer, surrounding observer, freely falling observer, and escaping observer, respectively.}
\label{TSPbh1}
\end{figure*}

We show the effect of the plasma on the shadow size in Fig.~\ref{TSP}, where we take $r_B=0.5r_S$ and $\beta=1$. When $r$ is large, the sizes of the black hole shadows for four different kinds of observers satisfy $\gamma_{\text{es}}>\gamma_{\text{st}}>\gamma_{\text{sur}}>\gamma_{\text{ff}}$. When $r$ is small, the shadow sizes for the escaping observer will smaller than that of for the static observer and surrounding observer, but it always larger than that of for the freely falling observer. But the existence of plasma makes the shadow size much smaller than the vacuum case for the surrounding observer and freely falling observer. In the presence of plasma, in some situation, the size of shadow does not monotonically decrease with $r$, (see the blue dashed line and purple dot dashed line in Fig.~\ref{TSP3os}). In Fig.~\ref{TSPbeta3os}, we show the effect of the parameter $\beta$ on the shadow size, where the observers locate at $r=3r_S$. For the escaping observer, the shadow size increases with the parameter $\beta$. For the other three kinds observers, the shadow size decreases with the parameter $\beta$. Note that, we only consider a specific plasma model and neglect the pressure of the plasma. More rich plasma models or dispersive medium models could be studied in future.

The reason for the nonmonotonically decreasing phenomenon is the combined result of $u_{\mu} k^{\mu}$ and $k_{\mu} w^{\mu}$. With the presence of plasma, this phenomenon occurs for the surrounding observer, freely falling observer and escaping observer. But this phenomenon occurs only when $r_B$ is greater than some values which depends on $\beta$ for the freely falling observer. Figure~\ref{TSPbh1} shows that, even $\beta=0.01$, we can also find the nonmonotonically decreasing phenomenon for the surrounding observer, freely falling observer, and escaping observer.

\begin{figure*}[htb]
\begin{center}
\subfigure[$ $]  {\label{RNnop}
\includegraphics[width=5.7cm]{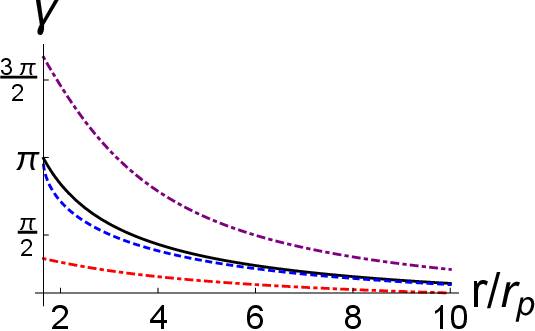}}
\subfigure[$ $]  {\label{RNwithpfour}
\includegraphics[width=5.7cm]{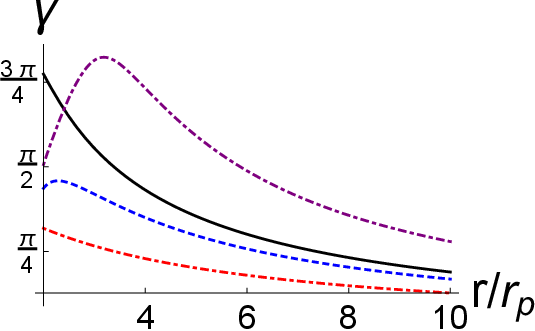}}
\end{center}
\caption{The angle $\gamma$ of the RN black hole as a function of the radial distance $r/r_p$. The inner horizon $r_{\text{m}}$ is set to $r_{\text{m}}=0.5r_{\text{p}}$. The black solid line, blue dashed line, red dot dashed line, and purple dot dashed line correspond to the static observer, surrounding observer, freely falling observer, and escaping observer, respectively. (a) Without plasma. (b) With plasma.}
\label{RN}
\end{figure*}

We compute the shadow size of the Reissner-Nordstr\"{o}m (RN) black hole by this method. We take the event horizon $r_{\text{p}}=M+\sqrt{M^2-Q^2}=1$, and the range of inner horizon $r_{\text{m}}=M-\sqrt{M^2-Q^2}$ is $[0,1]$. Some results are shown in Fig.~\ref{RN}. From Fig.~\ref{RNnop} we can see that, the relation of the angle $\gamma$ for the four kinds of observers is $\gamma_{\text{es}}>\gamma_{\text{st}}>\gamma_{\text{sur}}>\gamma_{\text{ff}}$ which is the same as the charged black hole with scalar hair case. And for the four kinds of observers, the angular diameters decrease with $r$. The nonmonotonically decreasing phenomenon does not occur. With the existence of plasma, the nonmonotonically decreasing phenomenon occurs for the surrounding observer (the blue dashed line in Fig.~\ref{RNwithpfour}) and the escaping observer (the purple dot dashed line in Fig.~\ref{RNwithpfour}), but it does not occur for the freely falling observer. Compared with these results, for the charged black hole with scalar hair, the nonmonotonically decreasing phenomenon occurs for the freely falling observer with and without plasma.

\section{Conclusions and discussions}\label{conclusion}

In this paper, we studied the shadow size of the static spherically symmetric four-dimensional charged black hole with scalar hair in terms of astrometrical observables. Using the Hamiltonian approach we derived the null geodesics. Based on the condition of the unstable circle orbit, we obtained the values of the conserved quantities $\kappa$ and $b$. And the photon sphere locates at $\fc{3}{2}r_S$ in the vacuum case which is similar to the Schwarzschild black hole. The shadow size of the charged black hole with scalar hair is determined by two light rays from the photon sphere with opposite angular momentum. Then we studied the shadow sizes for four kinds of observers, i.e., the static observer, surrounding observer, freely falling observer, and the escaping observer. We derived the angular diameter for these four kinds of observers in the vacuum background, respectively. We found that, at the same position the escaping observer will observe the largest shadow and the freely falling observer will observe the smallest one. The parameter $r_B$ of the charged black hole with scalar hair can also affect the shadow size. For the surrounding observer and freely falling observer: the larger the parameter $r_B$, the smaller the shadow size. For the escaping observer, the larger the parameter $r_B$, the larger the shadow size. Besides, for the freely falling observer, the shadow size does not decrease with $r$ monotonically in some cases.

The plasma as a dispersive medium can affect the trajectory of light rays. We got the four-momentum of light rays by making use of the Hamiltonian of light rays in the plasma. We took a spherically symmetric nonmagnetized pressureless neutral plasma as an example to study the effect of plasma. In this model, the plasma is consist of neutral hydrogens which are freely falling into the black hole. We numerically solved the radius of the photon sphere for different values of the parameters $r_B$ and $\beta$. For a smaller $r_B$, $r_{\text{sp}}$ increases with $\beta$, and for a larger $r_B$, $r_{\text{sp}}$ decreases with $\beta$, which was shown in Table~\ref{photonspher}. Although when $r$ is large, the shadow sizes for different observers also satisfy the relation $\gamma_{\text{es}}>\gamma_{\text{st}}>\gamma_{\text{sur}}>\gamma_{\text{ff}}$, the plasma makes the shadow sizes smaller than the vacuum case, which can be found in Fig.~\ref{TSP}. Especially, the nonmonotonically decreasing phenomenon occurs for the surrounding observer.

Compared with the RN black hole, for the freely falling observer, the nonmonotonically decreasing phenomenon occurs for both the case of with and without plasma. However, this only occurs when the observer is very close to the black hole, so it is almost impossible to detect it. Nevertheless, this is also helpful to understand the black hole. Besides, we only considered a simple plasma model, we should study more realistic models in future.

\section{Acknowledgments}

This work was supported by National Key Research and Development Program of China (Grant No. 2020YFC2201503), the National Natural Science Foundation of China (Grants No. 12147166, No. 11875151, No. 12075103, and No. 12247101), the China Postdoctoral Science Foundation (Grant No. 2021M701529), the 111 Project (Grant No. B20063), and Lanzhou City's scientific research funding subsidy to Lanzhou University.

\rule[-10pt]{14.3cm}{0.05em}\\
\\

\end{document}